# Lessons from the Failure and Subsequent Success of a Complex Healthcare Sector IT Project

David Greenwood, Ali Khajeh-Hosseini, Ian Sommerville

Dependable Socio-Technical Systems Engineering Group,
School of Computer Science, University of St Andrews, UK
`{dsg22, akh, ifs}@cs.st-andrews.ac.uk`

**Abstract.** This paper argues that IT failures diagnosed as errors at the technical or project management level are often mistakenly pointing to symptoms of failure rather than a project's underlying socio-complexity (complexity resulting from the interactions of people and groups) which is usually the actual source of failure. We propose a novel method that adopts a socio-complexity lens, Stakeholder Impact Analysis, that can be used to identify risks associated with socio-complexity as it is grounded in insights from the social sciences, psychology and management science. This paper demonstrates the effectiveness of Stakeholder Impact Analysis by using the 1992 London Ambulance Service Computer Aided Dispatch project as a case study, and shows that had our method been used to identify the risks and had they been mitigated, it would have reduced the risk of project failure. This paper's original contribution comprises adopting a socio-complexity lens and expanding upon existing accounts of failure by examining failures at a level of granularity not seen elsewhere that enables the underlying socio-complexity sources of risk to be identified.

## 1      Introduction

IT failures have been plaguing the implementation of IT systems since their introduction into organisations in the 1960s. Today's economic practices are more dependent upon IT than ever before and therefore understanding and preventing IT failures is even more worthy of academic study and industrial investment. There is a disciplinary history of analysing IT failures as technical failures and in more recent times broadening this to include project management and organisational factors. This paper's original contribution expands upon the existing body of IT failure studies by arguing that IT failures diagnosed as errors at the technical or project management level are often mistakenly pointing to symptoms of failure rather than a project's underlying socio-complexity which is regularly the actual source of failure.

This paper demonstrates the effectiveness of Stakeholder Impact Analysis by using the 1992/1996 London Ambulance Service Computer Aided Dispatch

(LASCAD92/96) projects as a case study, and shows that had our method been used to identify the risks and had they been mitigated, it would have reduced the risk of project failure. The Stakeholder Impact Analysis method is based on a framework that examines failures at a level of granularity not seen elsewhere, which enables the underlying socio-complexity sources of risk to be identified rather than its symptoms at the technical and project management level.

The paper is structured such that: Section 2 introduces the case-study, and studies of IT failure and their deficiencies; Section 3 describes socio-complexity and the Stakeholder Conflict Matrix; Section 4 details the paper's case-study methodology; Section 5 summarises the results; Section 6 describes the lessons learned; and Section 7 concludes the paper and discusses future work.

## 2  Background

### 2.1  Studies of IT Failure

Information System Development (ISD) failure is an ongoing theme in the study of large scale complex IT systems and is especially well researched in the Information Systems (IS) literature. Failure is, by nature, not a well-defined concept as it is the consequence of an evaluation that a stakeholder applies to a system, or project, at a particular time with respect to their expectations [1]. It has been demonstrated that over-time a stakeholder's perception of an ISD project changes with time and is dependent upon their perspective and the legitimacy of other voices over time [2]. In consequence the literature identifies many broad types of failure e.g. Correspondence failure; Process failure; and Interaction failure [3, 4]. Correspondence failure where the 'wrong' thing is delivered but to the 'correct' budget and schedule, process failure is where the 'correct' thing is delivered but to the wrong budget or schedule, and interaction failure where the 'correct' thing is delivered to the 'correct' budget and schedule but stakeholders do not use the system

ISD failure is studied using two dominant approaches, identifying failure factors, and identifying failure processes/dynamics [1]. Both approaches are based on the hypothesis that if the causes of failure can be identified, we can monitor and control those risks. Early studies focused on simple causes and often identified the cause as the shortcomings of individual IT practitioners or IT managers. However [5] identified that most of the difficulties were social and behavioural rather than technical. This result was confirmed and generalised by [6] and later by [7].

The bulk of the research has therefore identified social and behaviour factors associated with failure and these factors have typically been treated as causes of failure despite the fact that few large-scale statistical studies exist to illustrate how risk factors correlate with ISD project performance.

## 2.2 Risk Identification & Management Methods

Risks are events or situations that may adversely affect project performance. According to [8] risk exposure (RE) is a function of the probability of an adverse outcome P(OU) and the loss to parties if the adverse outcome occurs L(OU); therefore, RE = P(OU)*L(UO). Risk is managed in IT projects using a two-step procedure: risk assessment, and risk control. Risk assessment comprises identification, analysis and prioritization, which means possible sources of events that may cause adverse outcomes are elicited, their likelihood of occurrence and their consequences are estimated and then finally the risks are ranked for importance. Risk control comprises planning and management, resolution and monitoring, which means plans are made documenting how each risk will be dealt with (e.g. avoidance, transfer, reduction), these plans are executed and then the effectiveness of the plans are monitored and corrective action is taken where appropriate.

Project risks are normally identified by a risk analyst using their personal experience, project files, historical data and supplementary data collected on the basis of one-to-one interviews or analyst led working groups. In all these situations support tools such as checklists, models or prompts may be used to focus attention on a particular area of risk [9, 10]. There exist many checklists and models which focus on different aspects of projects [8, 11-14] e.g. aspects that authors believe significantly contribute to process failure, correspondence failure, or interaction failure. These identification methods are incorporated into boarder risk management approaches such as 'RiskMan', 'RISKIT' and SODIS [15-17]. We argue that the weakness that these existing approaches suffer from is that they point to symptoms of failure rather than a project's underlying socio-complexity (complexity resulting from the interactions of people and groups), which is usually the actual source of failure and thus the risk that needs to be addressed.

## 2.3 The Case study – The London Ambulance Service & LASCAD

The London Ambulance Service (LAS) is an NHS Trust that provides emergency ambulance service to the whole of the London area (approx 620 square miles). Similar to other public sector organisations, the LAS exhibits significant socio-complexity. In 1990 it was involved in a protracted pay and conditions dispute with its workers' union. In 1991, the 268 senior and middle management posts in the LAS were cut to 53. The official report of the enquiry into the LAS also commented that this restructuring caused a great deal of anxiety to workers creating the perception of continual pressure to down-size and improve performance [18].

The London Ambulance Service Computer Aided Despatch (LASCAD) project is an exemplar of an IT enabled work-transformation project. It comprised the automation of the dispatch of ambulances from call taking to ambulance dispatch. The need for the automation project was identified in the mid 1980s when the government perceived the London Ambulance Service to be failing to modernise and generally invest in its work force. After a failed modernisation attempt in 1987, a second attempt was initiated in October 1990. The planned implementation date was 8[th] of

January 1992 but by March 1992, the second phase of live trials was suspended due to the users of the system not having confidence in the system resulting in the Nation Union of Public Employees to get involved [8]. Until the 26$^{th}$ October 1992, the automated system struggled to satisfy its objectives resulting in ambulances being scheduled inefficiently and the system was finally switched off in the following 48 hours.

Following this failure a public enquiry was performed as the failure had become one of the highest profile IT failures in the UK. The project was revitalised by the newly appointed management after its failure in 92, however rather than pursuing the same approach as the LASCAD92 failure they opted for a radically different approach. The new project, called LASCAD96 in this paper, comprised a non-time pressured in-house development project. A custom off-the-shelf (COTS) solution was evaluated, but rejected, and a participative approach utilising prototyping was adopted to generate user participation and ownership [4]. The initial system was extremely simple and improvements were released in small increments where by September 1996 more radical enhancements were being accepted by the user-base resulting in a jump in productivity from 38%-60% of calls being despatched in 3minutes [4].

## 3      Socio-complexity and the Stakeholder Conflict Matrix

Complexity is the quality or property of being complicated such that an actor is unable to make predictions about the consequences of an action as the variables or their interactions are not well understood. Socio-complexity comprises the complicatedness associated with interactions between people and groups of people. This is observed in an organisation, or project organisation, in terms of pluralities of perspectives and pluralities of behaviour. Pluralities of perspective can be caused by ambiguity, uncertainty, un-surfaced assumptions, differences in interpretive norms, or group processes [19]. Pluralities of behaviour can be caused by differences in expectations, intentions and behavioural norms [19].

Stakeholder resistance is a form of conflict and as such is the embodiment of socio-complexity and a consequent of the tensions between the pluralities of stakeholders' perspectives and behaviours. Stakeholder resistance can be viewed as a feedback mechanism between stakeholders about the goodness of fit between their local environment and the intended project. It is for this reason that the Stakeholder Impact Analysis method appears to be an interesting risk analysis method as by understanding the kinds of thing that would make specific stakeholders, or stakeholder groups resist, many specific risks can be identified upfront in a systematic manner.

The Stakeholder Impact Analysis method used in this case study is an operationalisation of the Stakeholder Conflict Matrix [20]. The Stakeholder Conflict Matrix is the product of a multi-disciplinary literature survey of the sources of conflict in social settings. It comprised a review of over 50 articles in journals covering Applied Psychology, Administrative Science, Management Science, Social Science, Computer Science and Information Systems. The Stakeholder Conflict Matrix helps address the limitations of existing studies by approaching failure from a conflict perspective thus allowing rigorously designed studies from diverse fields to be brought to bear upon our understanding of failure thus providing much needed granularity and corroborating findings within CS/Informatics domain. The literature review findings are condensed into the Stakeholder Conflict Matrix (Table 1).

Table 1 - Stakeholder Conflict Matrix

| Conflict | Individual | Intra-group | Inter-group |
|---|---|---|---|
| Task | N/A | Disagreement over what to do to achieve aims | |
| Process | N/A | Disagreement over how to accomplish task | |
| Relational (Procedural / Distributive Injustice) | N/A | Unjust delegations of status, duty or resources **within** the group | Unjust delegations of status, duty or resources **between** groups |
| Role (Time, Resources and Capabilities) | Incompatibility between required activity and practical constraints | | |
| Role (Values, Status and Satisfaction) | Incompatibility between required activity and an **individual's** values, status or satisfaction | | Incompatibility between required activity and **group** values, status and satisfaction |
| Role (Multiple) | An **individual** is assigned multiple roles with incompatible activities or assessment metrics | | A **group** is assigned multiple roles with incompatible activities or assessment metrics |
| Relational (Incompatibility between actors) | N/A | Negative emotionality between **individuals** | Negative emotionality between **groups** |

## 4 Methodology

A case study methodology was adopted to evaluate the effectiveness of the Stakeholder Impact Analysis. Data on the LASCAD project was collected from five separate sources: [18] [21] [22] [23] [24] and it was verified that each account broadly corroborated one another other to ascertain the reliability of the data. The data was analysed using Stakeholder Impact Analysis (SIA) which is an operationalisation of the Stakeholder Conflict Matrix. SIA comprises: 1. Identifying key stakeholders; 2. Identifying changes in *what* tasks stakeholder would be required to perform and *how* they were to perform them; 3. Hypothesising what the likely consequences of the changes are with regards to stakeholders time, resources, capabilities, values, status and satisfaction; 4. Hypothesising the impact of these changes within the wider context of relational factors such as tense relationships between individuals or groups to which stakeholders belong; 5. Hypothesising whether the stakeholder will perceive the change as unjust (either procedurally or distributively) based upon the nature of changes and the stakeholders relational context.

The risks identified using Stakeholder Impact Analysis were used to test two hypotheses with the overall aim of illustrating that had Stakeholder Impact Analysis been performed during the LASCAD92 project and had the identified risks been mitigated it would have reduced the risk of project failure.

[H1] *The causes of the failure identified in other analyses map onto risks identified by the Stakeholder impact analysis*

If hypothesis one is supported by the results then this corroborates that Stakeholder Impact analysis captures the risks captured by existing approaches. The hypothesis was tested by identifying the causes of the LASCAD92 failure as identified by the Official Inquiry and supporting academic literature and for each cause identifying the relevant risks raised by the Stakeholder Impact Analysis.

[H2] *Risks identified in the failed LASCAD92 project were appropriately mitigated/partially mitigated in the successful LASCAD96 project*

If hypothesis two is supported by the results then this corroborates that the kinds of risks identified by Stakeholder Impact Analysis have a significant impact on project performance if left unmitigated. Hypothesis two was tested by identifying if the changes to practices identified by [4] in their case-study of the successful turn-around (LASCAD96) mitigated or partially mitigated each risk identified by Stakeholder Impact Analysis.

## 5   Results

A summary of the results is presented to highlight the most significant risks identified by Stakeholder Impact Analysis. The results support both [H1] and [H2]. Hypothesis 1 is supported by the fact that all identified causes of failure in the literature map to risks identified by Stakeholder Impact Analysis thus illustrating it is able to comprehensively identify risks (See http://www.cs.st-andrews.ac.uk/~dsg22/LASCAD/tables.pdf). Hypothesis 2 is supported by the fact that Stakeholder Impact Analysis found 24 stakeholder risks associated with the LASCAD 92 and 96 projects. Of these 24 risks, 2 were appropriately mitigated or partially mitigated during LASCAS 92. In contrast during LASCAD 96, 23 of the 24 risks were appropriately mitigated or partially mitigated (See table 2). Since LASCAD 96 was considered a success this suggests that the Stakeholder Impact Analysis identifies risk factors that if left unmitigated contribute to stakeholder resistance and ultimately project failure. These findings give grounds for the statement that had Stakeholder Impact Analysis been performed during LASCAD 92 it would have reduced the risk of project failure.

Table 2 - Mitigated vs unmitigated risks and project outcomes

| # | Partially Mitigated | Unmitigated | Project Outcome |
|---|---|---|---|
| LASCAD92 | 2 | 22 | Failure |
| LASCAD96 | 23 | 1 | Success |

Due to space restrictions only highlights of each stakeholders' perspective will be provided starting first with control room staff, then LAS management and finally with ambulance crew. It was identified that control room staff may oppose the implementation for the following reasons: [C5] the system could be perceived to reduce their status as it was designed to routinise their work removing many elements of skill/local knowledge that previously existed; [C7] the project was a top-down initiative and there was a history of conflict and non-cooperation with management; [C8] the project could be perceived as distributively unfair as staff received little benefit from the change but would need to retrain and there was a possibility of job cuts. In LASCAD96 the first risk [C5] was partially mitigated by the software being designed to support decision-making rather than automate it. This mitigated the risk as rather than routinising work and removing elements of skill it supplemented staff skill and knowledge. The second risk [C7] was partially mitigated by significant changes to LAS management and the building of cooperation by fulfilling staff requests e.g. the provision of a new more comfortable control room and the hiring of additional staff to reduce the burden of transition on existing staff. This mitigated the risk because management were less likely to be perceived as adversaries and partners to cooperate with. The third risk [C8] was mitigated as staff received improved working conditions and were given control of the roll out by being given a right to veto changes that they did not approve of. This mitigated the risk because staff now had control over changes and therefore were able to reject changes that were perceived to be deliver little net benefit.

It was identified that the LAS management may oppose the project for the following reasons: [M2] they perceive that they would be given inadequate time/resource to make the project a success; [M3] they do not have the will or ability to develop skills to manage the system or its development; [M5] may be reluctant to report negative information to executes for fear of losing their status or job; [M6] due to poor past relations management may view control room staff or ambulance crew as being obstructive when providing genuine negative feedback about the system. The first risk [M2] was mitigated in LASCAD92 by management being fearful for their jobs however the consequence of this was that management were reluctant to provide negative feedback (bad news) for fear of being seen as obstructive and thus losing their job. In LASCAD96 the risk was mitigated by an executive level commitment to provide whatever resources and time needed as well as fostering an atmosphere of openness. This mitigated the risk managers as the availability of resources and time reduced the perception of inadequate resources or time being made available. The second risk [M3] was mitigated in LASCAD92 by top-down pressure to meet ORCON targets and risks of job losses. In LASCAD96 this risk was mitigated by topdown pressure to meet targets, the provision of additional resources and restructuring to bolster management and operational staff. This mitigated the risk of lack of will or ability by providing a supportive atmosphere for staff and bringing in staff with additional skills and experience for others to learn from. The third risk [M5] was mitigated in LASCAD96 by introducing a flexible time-frame thus removing pressure for immediate results and also recent hiring of additional staff reinforcing the message that jobs were not at risk. This mitigated the risk because the reporting of negative information was no longer perceived as adversarial / a problem as there was plenty of time and there was no atmosphere of job losses. The forth risk [M6] was not mitigated in LASCAD92 or 96.

It was identified that ambulance crew may oppose the project for the following reasons: [A2] the system could be perceived to interfere with crew values of 'rapid response' if it does not take into account crew experience and local knowledge; [A7] the system could be perceived as management interference and procedurally unjust due to ongoing issues with staff consultation; [A8] the system could be perceived as distributively unjust are crew lose a their autonomy from which derive little benefit. The first and second risks [A2][A7] were mitigated in LASCAD96 by entering into consultation with staff and involving crew with testing and approval of equipment prior to go-live. This mitigated the risks as crews could influence changes to ensure they did not conflict with their values or modes of operating. The third risk [A8] was mitigated by improving crew working conditions by issuing them with personal radios so they could communicate with themselves and also upgrading ambulances for comfort. This mitigated the risk because the crews could shape the changes to ensure the distribution of benefit/loss was fair as they perceived it.

## 6   Lessons Learned

The LASCAD 92/96 project emphasises the importance of stakeholder resistance and its impact on the success of IT projects. The following list identifies the main lessons learned from this case-study:

- Resistance comprises stakeholders providing feedback on how they perceive a system to impact their local environment and therefore addressing their perceived risks is valuable as it facilitates good fit between their local and the system and the changes it brings about.
- The majority of stakeholder risks can be appropriately mitigated, or partially mitigated, by senior management demonstrating that they are willing to invest the resources it takes to get a project done well and that they are open and respond to continuous consultation/feedback from all stakeholders.
- Sources of continuous consultation/feedback are an important practice to mitigate risk these include: up-front consultation; ongoing drop-in sessions; user acceptance testing where users can delay go-live if unhappy.
- Software and system design should be mindful of stakeholder values, satisfaction, and status. One particularly important area is to avoid fully automating user decision-making and instead supporting the user to make better decisions. This is beneficial as it reduces the risk of removing satisfying or status granting aspects work whilst simultaneously enabling the user to incorporate their local knowledge or expertise.
- Stakeholder risks that are mitigated via coercion tend to dampen feedback loops between stakeholders resulting in poor communication and ultimately a project that is not a good fit with its environment and thus a failure.

# 7   Conclusion

It is concluded that: i) IT failures diagnosed as errors at the technical or project management levels are often mistakenly pointing to symptoms of failure rather than a project's underlying socio-complexity which is regularly the actual source failure; ii) had Stakeholder impact analysis been performed during the LASCAD92 project and the identified risks had been mitigated it would have reduced the risk of project failure.

Claim i) was corroborated by the fact that the LASCAD case study demonstrates that: a) most risks can be appropriately mitigated, or partially mitigated, by senior management demonstrating that they are willing to invest the resources it takes to get a project done well and that they are open and respond to continuous consultation/feedback from all stakeholders; b) sources of continuous consultation/feedback are an important practice to mitigate risks these include: up-front consultation; ongoing drop-in sessions; User acceptance testing where users can delay go-live if unhappy; c) risks that are mitigated via coercion tend to dampen feedback loops between stakeholders resulting in poor communication and ultimately a project that is not a good fit with its environment and thus a failure.

Claim ii) was corroborated by the fact that: d) the risks identified map onto causes of failure identified in other analyses of the failure; e) all the risks identified in the LASCAD92 project were mitigated in the successful LASCAD96 project suggesting their mitigation contributed to the success of the project. The conclusion is limited by the usual limitations of case-study research.

Future work is planned to corroborate these findings with other case studies and also perform statistical studies of IT project performance to establish the average performance impact of each kind of Socio-complexity derived risk factor as indicated in the Stakeholder Conflict Matrix.

# References

1.      Sauer, C.: Deciding the future for IS failures not the choice you might think. In: Currie, W.L., Galliers, B. (eds.): Rethinking Management Information Systems. Oxford University Press (1999) pp.279-309
2.      Wilson, M., Howcroft, D.: Re-conceptualising failure: social shaping meets IS research. Eur J Inf Syst **11** (2002) 236-250
3.      Lyytinen, K., Hirschheim, R.: Information systems failures - a survey and classification of the empirical literature. Oxford Surveys in Information Technology. Oxford University Press, Inc. (1987) 257-309
4.      Fitzgerald, G., Russo, N.L.: The turnaround of the London ambulance service computer-aided despatch system (LASCAD). Eur. J. Inf. Syst. **14** (2005) 244-257
5.      Colton, K.W.: Computers and Police: Patterns of Success and Failure. Sloan Management Review **2** (1972) 75-98
6.      Lucas, H.C.: Why information systems fail. Columbia University Press, New York (1975)
7.      Boland, R., Hirschheim, R.: Series Foreword. In: Boland, R., Hirschheim, R. (eds.): Critical Issues in Information Systems Research. Wiley, Chichester (1987)

8. Boehm, B.W.: Software Risk Management: Principles and Practices. IEEE Softw. **8** (1991) 32-41
9. Chapman, R.J.: The effectiveness of working group risk identification and assessment techniques. International Journal of Project Management **16** (1998) 333-343
10. Chapman, R.J.: The controlling influences on effective risk identification and assessment for construction design management. International Journal of Project Management **19** (2001) 147-160
11. Carr, M.J., Konda, S.L., Monarch, I., Ulrich, F.C., Walker, C.F.: Taxonomy-Based Risk Identification. (June 1993)
12. Moynihan, T.: How Experienced Project Managers Assess Risk. IEEE Softw. **14** (1997) 35-41
13. Keil, M., Cule, P.E., Lyytinen, K., Schmidt, R.C.: A framework for identifying software project risks. Commun. ACM **41** (1998) 76-83
14. Schmidt, R., Lyytinen, K., Keil, M., Cule, P.: Identifying Software Project Risks: An International Delphi Study. J. Manage. Inf. Syst. **17** (2001) 5-36
15. Carter, B.: Introducing Riskman: The European Project Risk Management Methodology. Stationery Office Books (1996)
16. Kontio, J.: The Riskit Method for Software Risk Management. Institute for Advanced Computer Studies and Department of Computer Science University of Maryland
17. Gotterbarn, D., Rogerson, S.: Responsible Risk Assessment with Software Development: Creating the Software Development Impact Statement. Communications of the Association for Information Systems **15** (2005)
18. Page, D., Williams, P., Boyd, D.: Report of the Inquiry Into The London Ambulance Service. The Communications Directorate, South West Thames Regional Health Authority, London (1993)
19. Greenwood, D.: A Normative Agent Organisational Modelling approach to aid the analysis of collaborative business processes spanning multiple enterprises: A modelling approach to aid Service Oriented Information System development in business organisations.: Informatics, Vol. MSc. University of Reading, Reading (2007) 97
20. Greenwood, D.: A Literature Survey of Unsupportive Stakeholder behaviour and its impact on the development and deployment of Complex IT systems. University of St Andrews, St Andrews (2010) 29
21. Beynon-Davies, P.: Information systems `failure': the case of the London Ambulance Service's Computer Aided Despatch project. Eur J Inf Syst **4** (1995) 171-184
22. Finkelstein, A., Dowell, J.: A comedy of errors: the London Ambulance Service case study. Proceedings of the 8th International Workshop on Software Specification and Design. IEEE Computer Society (1996)
23. Hougham, M.: London Ambulance Service computer-aided despatch system. International Journal of Project Management **14** (1996) 103-110
24. Beynon-Davies, P.: Human error and information systems failure: the case of the London ambulance service computer-aided despatch system project. Interacting with Computers **11** (1999) 699-720